\documentclass[12pt,epsf]{article}
\usepackage{graphicx,epsfig,amsmath,amssymb}
\begin{document}

\def\a{\alpha}
\def\b{\beta}
\def\c{\varepsilon}
\def\d{\delta}
\def\e{\epsilon}
\def\f{\phi}
\def\g{\gamma}
\def\h{\theta}
\def\k{\kappa}
\def\l{\lambda}
\def\m{\mu}
\def\n{\nu}
\def\p{\psi}
\def\q{\partial}
\def\r{\rho}
\def\s{\sigma}
\def\t{\tau}
\def\u{\upsilon}
\def\v{\varphi}
\def\w{\omega}
\def\x{\xi}
\def\y{\eta}
\def\z{\zeta}
\def\D{\Delta}
\def\G{\Gamma}
\def\H{\Theta}
\def\L{\Lambda}
\def\F{\Phi}
\def\P{\Psi}
\def\S{\Sigma}

\def\o{\over}
\def\beq{\begin{eqnarray}}
\def\eeq{\end{eqnarray}}
\newcommand{\gsim}{ \mathop{}_{\textstyle \sim}^{\textstyle >} }
\newcommand{\lsim}{ \mathop{}_{\textstyle \sim}^{\textstyle <} }
\newcommand{\vev}[1]{ \left\langle {#1} \right\rangle }
\newcommand{\bra}[1]{ \langle {#1} | }
\newcommand{\ket}[1]{ | {#1} \rangle }
\newcommand{\EV}{ {\rm eV} }
\newcommand{\KEV}{ {\rm keV} }
\newcommand{\MEV}{ {\rm MeV} }
\newcommand{\GEV}{ {\rm GeV} }
\newcommand{\TEV}{ {\rm TeV} }
\def\diag{\mathop{\rm diag}\nolimits}
\def\Spin{\mathop{\rm Spin}}
\def\SO{\mathop{\rm SO}}
\def\O{\mathop{\rm O}}
\def\SU{\mathop{\rm SU}}
\def\U{\mathop{\rm U}}
\def\Sp{\mathop{\rm Sp}}
\def\SL{\mathop{\rm SL}}
\def\tr{\mathop{\rm tr}}

\def\IJMP{Int.~J.~Mod.~Phys. }
\def\MPL{Mod.~Phys.~Lett. }
\def\NP{Nucl.~Phys. }
\def\PL{Phys.~Lett. }
\def\PR{Phys.~Rev. }
\def\PRL{Phys.~Rev.~Lett. }
\def\PTP{Prog.~Theor.~Phys. }
\def\ZP{Z.~Phys. }

\newcommand{\bea}{\begin{eqnarray}}
\newcommand{\eea}{\end{eqnarray}}
\newcommand{\bear}{\begin{array}}
\newcommand {\eear}{\end{array}}
\newcommand{\bef}{\begin{figure}}
\newcommand {\eef}{\end{figure}}
\newcommand{\bec}{\begin{center}}
\newcommand {\eec}{\end{center}}
\newcommand{\non}{\nonumber}
\newcommand {\eqn}[1]{\beq {#1}\eeq}
\newcommand{\la}{\left\langle}
\newcommand{\ra}{\right\rangle}
\newcommand{\ds}{\displaystyle}
\newcommand{\hyp}{\,{\rm \mathchar`-}\,}
\def\SEC#1{Sec.~\ref{#1}}
\def\FIG#1{Fig.~\ref{#1}}
\def\EQ#1{Eq.~(\ref{#1})}
\def\EQS#1{Eqs.~(\ref{#1})}
\def\TEV#1{10^{#1}{\rm\,TeV}}
\def\GEV#1{10^{#1}{\rm\,GeV}}
\def\MEV#1{10^{#1}{\rm\,MeV}}
\def\KEV#1{10^{#1}{\rm\,keV}}
\def\lrf#1#2{ \left(\frac{#1}{#2}\right)}
\def\lrfp#1#2#3{ \left(\frac{#1}{#2} \right)^{#3}}
\def\oten#1{ {\mathcal O}(10^{#1})}


\baselineskip 0.7cm

\begin{titlepage}

\begin{flushright}
RIKEN-MP-59 \\
TU-923
\end{flushright}

\vskip 1.35cm
\begin{center}
{\large \bf
    Hybrid inflation in high-scale supersymmetry
    }
\vskip 1.2cm

\renewcommand{\thefootnote}{\fnsymbol{footnote}}

    Tetsutaro Higaki$^{\,a \star}$\footnote[0]{$^\star$ email: tetsutaro.higaki@riken.jp},
    Kwang Sik Jeong$^{\, b\ast}$\footnote[0]{$^\ast$ email: ksjeong@tuhep.phys.tohoku.ac.jp},
    Fuminobu Takahashi$^{\,b \dagger} $\footnote[0]{$^\dagger$ email: fumi@tuhep.phys.tohoku.ac.jp}

    \vskip 0.4cm
    {\it $^a$ Mathematical Physics Lab., RIKEN Nishina Center, Saitama 351-0198, Japan}
    \vskip 0.3cm
{\it $^b$ Department of Physics, Tohoku University, Sendai 980-8578, Japan}\\

\vskip 1.5cm

\abstract{
In hybrid inflation, the inflaton generically has a tadpole
due to gravitational effects in supergravity, which significantly changes the inflaton
dynamics in high-scale supersymmetry. We point out that the tadpole can be cancelled
if there is a supersymmetry breaking singlet with gravitational couplings, and
in particular, the cancellation is automatic  in no-scale supergravity. We consider
the LARGE volume scenario as a concrete example and discuss the compatibility between
the hybrid inflation and the moduli stabilization. We also point out that the dark radiation
generated by the overall volume modulus decay naturally relaxes a tension between the
observed spectral index and the prediction of the hybrid inflation.
}
\end{center}

\end{titlepage}

\setcounter{page}{2}

\section{Introduction}
\label{sec:1}

The recent observations of the cosmic microwave background radiation (CMB)~\cite{Komatsu:2010fb} strongly
suggest the inflationary era~\cite{Guth:1980zm} in the early Universe; the CMB temperature fluctuations extending
beyond the horizon scale at the last scattering surface can be interpreted as the evidence for
the accelerated expansion in the past.

The study of the CMB temperature anisotropies as well as the large-scale structure
provide us with a clue about the mechanism that laid  down the primordial density fluctuations.
Whereas some models are already excluded or strongly disfavored,  it is not enough at present to pin
down the mechanism.  From a minimalist point of view,
the quantum fluctuations of the inflaton are the most plausible candidate for the origin of
density fluctuations.

Among many inflation models proposed so far,  a hybrid inflation scenario~\cite{Copeland:1994vg} is of particular interest,
especially in the supersymmetric framework.
The inflationary trajectory is along the $F$-flat direction, and the inflation ends when
waterfall fields become tachyonic and develop a large vacuum expectation value (VEV).
The hybrid inflation has been extensively studied from various points of view~\cite{Lazarides:1991wu,Asaka:1999yd, Buchmuller:2000zm,
Nakayama:2010xf,Buchmuller:2012ex}.

It was pointed out in  Refs.~\cite{Buchmuller:2000zm, Nakayama:2010xf} that the inflaton has a tadpole in proportional
to the gravitino mass, because of the gravitational interaction in supergravity.  In particular, the inflaton dynamics is significantly affected
by the tadpole for
the gravitino mass $m_{3/2} \gtrsim {\cal O}(10^2)$\,TeV and the successful inflation is possible only in a very
tight corner of the parameter space~\cite{Nakayama:2010xf}.
On the other hand,
the recent discovery of the standard model (SM)-like Higgs boson with mass about
$125$  $\,{\rm \mathchar`-}\,$ $126$\,GeV~\cite{:2012gk,:2012gu}
may imply high-scale supersymmetry (SUSY)~\cite{Okada:1990gg,Giudice:2011cg}.
If so, there will be  a tension between the hybrid inflation and high-scale SUSY.

In this letter we point out that the tadpole can be cancelled if there is
a SUSY breaking singlet $Z$ with gravitational couplings, which is usually not required in high-scale SUSY because
the gaugino masses can be generated by the anomaly mediation~\cite{Giudice:1998xp,Randall:1998uk,Bagger:1999rd}.
 In the presence of such a singlet, the gravitinos are generically produced
by the inflaton decay~\cite{Kawasaki:2006gs,Dine:2006ii,Endo:2006tf,Asaka:2006bv,Endo:2006qk,Endo:2007ih,Endo:2007sz},
and moreover, coherent oscillations of the lowest component of $Z$
are produced after inflation~\cite{Coughlan:1983ci} and mainly decay into gravitinos. Thus, the Universe likely becomes gravitino-rich, which has
various interesting implications~\cite{Jeong:2012en}.
We will also see that  the cancellation of the tadpole is automatic in the no-scale supergravity~\cite{Cremmer:1983bf,Ellis:1983sf}.
However, the moduli stabilization is one of the central issue in the no-scale supergravity.
Lastly we consider a LARGE volume scenario (LVS)~\cite{Balasubramanian:2005zx},
as a concrete and realistic model where all the moduli are stabilized successfully, in order
to see whether the hybrid inflation can be successfully implemented.
We will also point out that the dark radiation
generated by the overall volume modulus decay~\cite{Higaki:2012ba,Cicoli:2012aq,Higaki:2012ar}
naturally relaxes a tension between the observed spectral index and the prediction of the hybrid inflation.

\section{Tadpole problem in hybrid inflation}
\label{sec:2}
In this section we briefly review the tadpole problem in the hybrid
inflation~\cite{Buchmuller:2000zm, Nakayama:2010xf}.
To see the essential features, we focus on the minimal supergravity model
described by
\bea
K &=& |\Phi|^2 + |\Psi|^2 + |\bar\Psi|^2 + |Z|^2,
\nonumber \\
W &=& \kappa \Phi(M^2-\Psi\bar\Psi) + W_0(Z),
\label{KW}
\eea
for the inflaton $\Phi$, and the waterfall fields $\Psi+\bar\Psi$
vector-like under a U(1) gauge symmetry, with the $R$ charges assigned
by $R(\Phi)=2$ and $R(\Psi)=R(\bar\Psi)=0$.
Here $Z$ is a SUSY breaking field, which we assume to have an $F$-term VEV
to cancel the vacuum energy density, without specifying the detailed
mechanism of stabilization.
We also assume that the inflaton is charged under an additional global
U(1)$_\Phi$ symmetry which is explicitly broken only by the $M^2$ term.
This approximate symmetry explains the hierarchy between the inflation
scale and the Planck scale ($M_P=1$).
We take $M$ and $\langle W_0 \rangle$ to be real and positive, which is
always possible through appropriate U(1)$_\Phi$ and U(1)$_R$ transformations.
Now the gravitino mass reads
\bea
m_{3/2} = \langle e^{K/2}W \rangle \simeq \langle W_0 \rangle,
\eea
at the vacuum lying near the $F$-flat direction $\Psi\bar\Psi=M^2$,
and thus is determined by $W_0$, which breaks U(1)$_R$ down
to a $Z_2$ subgroup.
Here one should be careful that $\langle e^{K/2} \rangle$ can be hierarchically
small in the LVS as we shall see later.

For a sufficiently large field value of $|\Phi|$ ($\gg M$), the waterfall
fields get large supersymmetric masses $\kappa|\Phi|$ and are
stabilized at the origin, $\Psi=\bar\Psi=0$.
Then integrating out the waterfall fields, one obtains
\bea
\Delta K_{\rm eff} = 
-\frac{\kappa^2}{16\pi^2}\ln\left(\frac{|\Phi|^2}{\Lambda^2}\right)|\Phi|^2,
\eea
which gives an important contribution to the scalar potential because
the $F$-term of $\Phi$ includes $\kappa M^2$ induced by
the linear superpotential term.
The inflaton scalar potential is written
\bea
V &=& \kappa^2 M^4
+ \frac{\kappa^4}{16\pi^2}M^4\ln\left(\frac{|\Phi|^2}{\Lambda^2}\right)
-2 m_{3/2} \kappa M^2 (\Phi+\Phi^*)
\nonumber \\
&&
+\, m^2_\Phi |\Phi|^2
+ \frac{1}{2}\kappa^2 M^4 |\Phi|^4
+ \cdots,
\eea
where $m^2_\Phi={\cal O}(m^2_{3/2})$, and the ellipsis denotes higher order
terms of $\Phi$ suppressed by $M_P$.
The hybrid inflation is implemented by the first two terms in the potential.
The inflaton $\Phi$ rolls down the (approximately) flat potential until
the waterfall fields become tachyonic at $|\Phi|\approx M$ and end
the inflation.
Indeed, in the absence of the tadpole term, the successful inflation is
achievable for a wide range of parameters,
$10^{-7}\lesssim \kappa \lesssim 10^{-1}$ and
$3\times 10^{14}\,{\rm GeV}\lesssim M \lesssim 10^{16}\,{\rm GeV}$,
where we have imposed the WMAP normalization on the
density perturbation~\cite{Komatsu:2010fb} and the cosmic string
constraint~\cite{Battye:2006pk}.

However the hybrid inflation can be spoiled by the tadpole term of the inflaton
\cite{Nakayama:2010xf}.
It is generated proportional to the gravitino mass,
more precisely to the $F$-term of the supergravity multiplet,
independently of the details of supersymmetry breaking.
This is because the $M^2$ term breaks the conformal symmetry explicitly.
Let us see the difficulties caused by the tadpole.
First, it generates an unwanted minimum at a large field value of $\Phi$,
and thus once the inflaton is trapped in the wrong vacuum, the inflation will
never end.
This problem can be avoided by fine-tuning the initial phase of the inflaton.
However, even in this case, a large tadpole makes the inflaton potential so steep
that the duration of the inflation becomes shorter, requiring the inflaton
to initially sit at larger field values.
Then, in order to satisfy the WMAP normalization, the inflation scale must be
higher,\footnote{Note that the density perturbation $\zeta$ scales as $|V^{3/2}/V'|$,
and the tadpole increases $|V'|$.
Here the prime denotes the differentiation with respect to the inflaton.}
which is constrained by the observational upper bound on the tension of the cosmic
string formed after the inflation ends.
These imply that the tadpole is dangerous for the hybrid inflation, especially
in the high-scale SUSY scenario where the gravitino mass is around or above
$10^2$ TeV.

To implement hybrid inflation successfully, one may add the inflaton coupling to
the SUSY breaking field,
\bea
\Delta K =  a Z + b Z |\Phi|^2 + {\rm h.c.},
\eea
with constants $a$ and $b$ for $\langle|Z|\rangle \ll 1$.
Note that the $aZ$ term contributes to the inflaton coupling since the supergravity
action depends on the K\"ahler potential through $-3e^{-K/3}$,
and also modifies the $F$-term of supergravity multiplet.
Thus the potential includes
\bea
V|_{\rm tadpole} = -\Big(2 - \sqrt3 (a-b)\Big) m_{3/2} \kappa M^2 (\Phi+\Phi^*),
\eea
where we have used $F^Z=-\sqrt3m_{3/2}M_P$,
and assumed the approximate U(1)$_\Phi$ symmetry.
The above tells us that a cancellation between tadpoles for the inflaton
occurs if one takes $a$ and $b$ to be
\bea
a-b = \frac{2}{\sqrt3},
\eea
which is nothing but fine-tuning, but there may be anthropic selection
of the parameters for the successful inflation.
Another way to avoid the tadpole problem, which would be more plausible,
is to consider hybrid inflation within no-scale supergravity.
Then, since the $F$-term of supergravity multiplet vanishes, no tadpoles
are induced even though the $M^2$ term breaks the conformal symmetry
explicitly.
We will examine this possibility in more detail in the next section.

Lastly we mention the cosmological aspects of such SUSY breaking singlet with Planck-suppressed couplings.
First, if $Z$ is stabilized during inflation at a place deviated from the low-energy minimum,
it will start to oscillate after inflation~\cite{Coughlan:1983ci}. The coherent oscillations of $Z$ may or may not dominate
the Universe, depending on the initial amplitude. If kinematically allowed, $Z$ generically
decays into gravitinos at a sizable rate.  Second, we expect that there are generically following interactions,
\bea
K & \supset &  |\Psi|^2 ZZ +  |{\bar \Psi}|^2 ZZ + {\rm h.c.} ,
\eea
which cause the gravitino production from the decay of the waterfall fields~\cite{Kawasaki:2006gs,Dine:2006ii,Endo:2006tf}.
Note that the energy of the Universe after inflation is dominated by the inflaton and the waterfall fields. In fact, the inflaton and
the waterfall fields are mixed with each other
due to the superpotential term $W_0$~\cite{Kawasaki:2006gs}.  Thus, the Universe likely becomes gravitino-rich
in the presence of such SUSY breaking singlet~\cite{Jeong:2012en}. The gravitino-rich Universe has various interesting
implications; the Wino-like LSP can account for dark matter if its mass happens to be of ${\cal O}(100)$\,GeV, even though
the other superparticles are much heavier. The singlet also makes the gravitino-induced baryogenesis~\cite{Cline:1990bw} possible,
if the R-parity is largely violated.

It is interesting that the SUSY breaking singlet $Z$ with gravitational couplings
is required to cancel the tadpole of the inflaton, which enables the hybrid inflation in high-scale SUSY,
while such singlet is not necessary from the phenomenological point of view because
of the anomaly mediation contribution to the gaugino mass.

\section{No-scale supergravity and LARGE Volume Scenario}
\label{sec:3}
In this section we consider  concrete examples where the tadpole
of the inflaton is naturally canceled or suppressed. First we consider
the no-scale supergravity, and then move on to the LVS which has
an approximate no-scale structure while all the modulus fields are stabilized.

Let us consider a no-scale model with a K\"ahler potential of the
form~\cite{Cremmer:1983bf,Ellis:1983sf}
\bea
K &=& -3 \ln\left(
T+T^\dag -\frac{1}{3} \sum_{i} 
|\phi_i|^2 \right),
\label{eq:ns}
\eea
and the superpotential
\bea
W_{\rm inf} = \kappa \Phi(M^2-\Psi\bar\Psi) + \omega_0,
\eea
for a SUSY breaking field $T$, where $\phi_i$ denotes the inflaton $\Phi$
and the waterfall fields $\Psi$ and $\bar \Psi$, and we have included the
constant superpotential.
As the following argument does not depend on the the waterfall fields that are
heavy during inflation, we will neglect them.
Note that the superpotential does not contain $T$, and thus the tree-level
potential for $T$ remains flat at the minimum, which is one of the notable
features in a no-scale supergravity model.

%

For the above no-scale model, one finds
\bea
V|_{\rm tadpole} = -2
\left( 1-\frac{1}{3}K^{I\bar J}K_IK_{\bar J} \right) m_{3/2}
\kappa M^2 (\Phi+\Phi^*) = 0,
\eea
where $m_{3/2}\simeq e^{K/2}\omega_0$.
Thus the tadpole problem is absent in the no-scale supergravity.
In fact, it is well known that the scalar potential in a no-scale supergravity model
resembles that in the global SUSY.
Note however that the modulus $T$ has a run-away potential during inflation, and therefore
it must be stabilized by modifying the K\"ahler potential~\cite{Ellis:1984bs} or adding the non-perturbative
effects~\cite{Gelmini:1983ji} or the $D$-term~\cite{Ellis:2006ara}.
It is then important to see whether the modulus stabilization revives
the tadpole problem or not.


In order to see if the hybrid inflation can be implemented together with the successful moduli stabilization of $T$,
let us study a LVS model based on the type IIB orientifold compactifications in flux vacua \cite{Balasubramanian:2005zx}
as a more realistic example.
We consider the model with relevant three K\"aher moduli on a Calabi-Yau (CY) space
with a singularity
\cite{Blumenhagen:2009gk}
\bea
\nonumber
K &=& -2\ln \bigg( {\cal V} + \frac{\xi}{2}\bigg) + \frac{(T_v+T_v^{\dag} + V_{U(1)})^2}{{\cal V}}
+  \frac{k_{\rm inf} }{{\cal V}^{2/3}}\bigg(
1 - \frac{\delta}{{\cal V}}
\bigg) K_{\rm inf}, \\
W &=& W_{\rm inf} + Ae^{-a T_s}  .
\eea
Here the three complexified K\"ahler moduli
$T_i = \tau_i + i \sigma_i ~(i=b,s,v)$ describe the overall 4-cycle volume,
the local 4-cycle volume, the singularity on the CY space respectively for the real parts $\tau_i$, while
the imaginary parts $\sigma_i$ are given by the integrands of the RR 4-form potentials on the corresponding cycles.
The CY volume ${\cal V}$ and the $\alpha'$-correction $\xi$ \cite{Becker:2002nn} are respectively given by
\bea
{\cal V} = (T_b +T_b)^{3/2} - (T_s + T_s^{\dag})^{3/2},
\qquad \xi = - \frac{ \chi({\rm CY})\zeta(3)}{2(2\pi)^3g_s^{3/2}} ,
\eea
where we will assume that $g_s = {\cal O}(0.1)$
is the string coupling and  $\chi ({\rm CY}) = 2(h^{1,1}({\rm CY})-h^{2,1}({\rm CY})) = -{\cal O}(100) < 0$
is the Euler number on the CY; one then finds that $\xi = {\cal O}(1)$.
For the inflaton sector, $k_{\rm inf} = 1 + \sum_{n \geq 1} b_n (T_v + T_v^{\dag})^n$ is $T_v$-dependent part with the Taylor expansion.
$K_{\rm inf}$ and $W_{\rm inf}$ are given by Eq.(\ref{KW}),
involving the constant superpotential $\omega_0 ={\cal O}(1)$
which comes from the 3-form fluxes.\footnote{
Although the tadpole $M^2$ of the inflaton in the superpotential can be also written by moduli
we will treat it as the constant in this paper,
because the tadpole problem will not become better by such moduli without a fine-tuning.
}
We again assume the presence of U(1)$_{\Phi}$ to simplify the following discussion.\footnote{
Such an U(1)$_{\Phi}$ should be a global one, otherwise
one will obtain the higher order terms of $\Phi$ from the D-term potential.
Therefore the U(1)$_{\Phi}$ might be just accidental or a symmetry in a local geometry on a CY space.
}
We have included the expected $\alpha'$-correction  parameterized by $\delta$
in the inflaton K\"ahler potential~\cite{Blumenhagen:2009gk},
because the matter wave function on the local model is not known well.
The non-perturbative term $Ae^{-a T_s}$ is considered as
instanton effect or gaugino condensation which arises from the E3-brane or D7-branes
wrapping on the local cycle supported by $\tau_s$.
Then one finds that $a = 2\pi/N$ and $N$ is a natural number while $A$ is expected to be order unity.

In this setup, both the visible and inflaton sectors are supposed to be realized on the singularity
supported by $T_v$,
and therefore we will have the anomalous U(1) vector multiplet $V_{U(1)}$ due to the chiral fermions.
Then $T_v$ has a supersymmetric St\"uckelberg coupling associated with $V_{U(1)}$ for the cancellation of the anomaly via
the Green-Schwarz mechanism that $T_v$ shifts under the U(1) transformation.
Hence $T_v$ becomes massive,
being absorbed into the gauge multiplet:
$m_{T_v} = m_{V_{U(1)}} \sim 1/{\cal V}^{1/2} = M_{\rm string}$.
Then one also finds that $\langle T_v \rangle = 0$ through the D-term potential
$D_{U(1)} \propto \partial_{T_v} K \propto \tau_v = 0.$

After integrating out the massive gauge multiplet which consists of $T_v$ and $V_{U(1)}$,
the simplified action is obtained,\footnote{
In this paper, we will not consider the quantum effects associated with anomalous $U(1)$ \cite{Shin:2011uk}
in addition to moduli-redefinitions \cite{Conlon:2010ji, Choi:2010gm} for simplicity.
However, it will be sufficient for the study of the tadpole problem.
Even if included, the situation is not improved because the no-scale structure is broken not only by the $\alpha'$-correction
but also by the quantum correction.
}
\bea
\nonumber
K &=& -2\ln \bigg( {\cal V} + \frac{\xi}{2}\bigg)
+  \frac{\tilde{K}_{\rm inf}}{{\cal V}^{2/3}}\bigg(
1 - \frac{\delta}{{\cal V}} \bigg), \\
W &=& W_{\rm inf} + Ae^{-a T_s}  .
\eea
Here note that $K_{\rm inf}$ can be modified to ${\tilde K}_{\rm inf}$; while
the inflaton $\Phi$ is assumed to be neutral under the anomalous U(1) and hence
the $\Phi$ part is not modified,
$\Psi + \bar{\Psi}$ can have quartic terms in the K\"ahler potential if they are charged.
For the systematic study of the tadpole of $\Phi$ during inflation,
let us rewrite the above effective action, neglecting $\Psi + \bar{\Psi}$:
\bea
\nonumber
K &=& -3 \ln \bigg(T_b + T_b^{\dag} - \frac{c_1}{3}|\Phi|^2 \bigg)
+ \frac{f \bigg(T_s + T_s^{\dag} - \frac{c_3}{3}|\Phi|^2 \bigg)}{(T_b+T_b^{\dag} - \frac{c_2}{3}|\Phi|^2)^n } , \\
W &=& \omega_0 + Ae^{-a T_s} + \kappa M^2 \Phi,
\label{KWeff}
\eea
for a positive rational number $n$.
Note that the presence of $c_2$ implies an $\alpha'$-correction in the K\"ahler metric of the inflaton.
Under the choice of $(n, c_1,c_2,c_3) =(3/2, 1, 2/3, 0)$ and $f(x) = 2 x^{3/2} -\xi $,
the above LVS model is reproduced and one then finds $\delta = \xi/3$;
this is the sequestered case in which
the inflaton K\"ahler potential is approximately given by $e^{K_{\rm moduli}/3}$ \cite{Blumenhagen:2009gk}.

Using the effective action Eq.(\ref{KWeff}),
the moduli are stabilized and the tadpole of $\Phi$ is then estimated as
\bea
V|_{\rm tadpole} &=&
\Big[
G_{\Phi} e^{G}(G_i G^i -3 ) + e^{G}( G^i \nabla_{\Phi} G_i + G_{\Phi})
\Big]_{\Phi =0} (\Phi+\Phi^*)
\label{tadpoleVSUGRA}
\nonumber \\
&=& r e^{K/2}m_{3/2} \kappa M^2 (\Phi+\Phi^*),
\eea
with
\bea
m_{3/2} &=& e^{K/2}W \sim \frac{\omega_0}{{\cal V}},
\nonumber \\
\label{tadpoler}
r &\simeq&
\frac{
-\frac{n(n-3)(2n-3)x}{ax}\,c_1
-\frac{n^2(n-3)x}{ax}\,c_2
+ t(n-1)(n+3)x c_3
}
{3(n+3) t^n c_1 +  \frac{2n^2 xf'}{ax} c_2- t  (n+3)f' c_3}
f^\prime,
\eea
where we have neglected small terms suppressed by $1/t^n$ in $r$,
and defined $t \equiv T_b +T_b^{\dag}$, $x \equiv T_s +T_s^{\dag}$,
and $f' = \partial_x f$.
See the appendix \ref{tadpolederi} for the derivation.
For $n=3/2$, the first term in the numerator in Eq.(\ref{tadpoler})
vanishes at the leading order of CY volume expansion,
while the second and third terms do not.
The third term will vanish when the inflaton is located on the singularity; $c_3 =0$.
However, it is expected that
we will always have non-zero $c_2$ due to the $\alpha'$-correction, especially in the inflaton K\"ahler potential.
In the original LVS model, one then finds
\bea
r & \simeq & \frac{c_2 x f'}{c_1 ax t^{3/2}} \sim
\frac{1}{{\cal V}\ln{\cal V}}, \\
V|_{\rm tadpole} & =& r e^{K/2}m_{3/2} \kappa M^2 (\Phi+\Phi^*)
\sim \frac{1}{{\cal V}^2\ln{\cal V}}\, m_{3/2} \kappa M^2 (\Phi+\Phi^*),
\label{LVStadpole}
\eea
while the vacuum energy during inflation is given by the no-scale one,
$V|_{\rm inf} = \kappa^2 M^4/{\cal V}^{4/3}$ at the tree-level.
Here we have used $x f' \sim x^{3/2} \sim \xi ={\cal O}(1)$ and
$a x \sim \ln {\cal V}$ obtained from the stationary conditions given in the appendix.

However, this is not the end of the story.
The LVS model has a negative cosmological constant
$\langle V_{\rm AdS} \rangle \sim -m_{3/2}^2/({\cal V}\ln{\cal V})$
at the  SUSY breaking minimum.
Therefore the uplifting is required for obtaining de Sitter/Minkowski vacuum \cite{Cicoli:2012fh}
as in the KKLT case \cite{Kachru:2003aw, Saltman:2004sn, Burgess:2003ic}.
As can be seen in Eq.(\ref{tadpoleVSUGRA}), there is an additional contribution from
the uplifting sector to the tadpole
of ${\cal O}({\cal V}^{-3})$,  because the sector contributes to the energy density of
$|\langle V_{\rm AdS} \rangle|$ in the scalar potential.
In particular, $G^i \nabla_{\Phi} G_i$ in Eq.(\ref{tadpoleVSUGRA}) depends on the coupling between the inflaton and the uplifting sector,
and therefore is quite model-dependent, and so we are left with the tadpole of order ${\cal V}^{-3}$ unless fine-tuning between
$\alpha'$-correction and the uplifting sector is assumed.
(Furthermore, if $M$ originates from a heavy modulus, another contribution to the tadpole could be present.)

To summarize, for the canonically normalized inflaton field
$\hat{\Phi} = \Phi /{\cal V}^{1/3}$, we generically have
\bea
V|_{\rm inflaton} = {\kappa}^2 \hat{M}^4 + \frac{\zeta }{{\cal V}}\,
m_{3/2} \kappa \hat{M}^2  (\hat{\Phi}+\hat{\Phi}^*)+\cdots,
\eea
at the tree-level.
Here $\hat{M} = M/{\cal V}^{1/3}$, and
$\zeta$ varies from $1/\ln {\cal V}$ to order unity, depending on the details of the
uplifting sector.
Note that the Yukawa coupling between the inflaton and waterfall fields does not depend on
${\cal V}$ in the canonical basis.
As expected from the fact that the approximate no-scale structure is broken at ${\cal V}^{-1}$,
the tadpole suppressed by ${\cal V}$ appears. Since ${\cal V}$ is large, the tadpole
problem is greatly relaxed in LVS, compared to the general case with the same gravitino mass.


Let us now discuss if the hybrid inflation works successfully in this framework with the suppressed
tadpole.
In principle the tadpole can be vanishingly small for sufficiently large ${\cal V}$. However,
${\cal V}$ cannot be too large,  because the modulus ${\rm Re}(T_b)$ would be extremely light,
causing several difficulties.
Note that the modulus obtains mass 
\beq
m_{T_b} \sim \frac{m_{3/2}}{{\cal V}^{1/2}\sqrt{\ln{\cal V}}} .
\eeq
In fact, the most important one is that from the modulus
destabilization. Through the uplifting of the AdS to a Minkowski vacuum,
the potential barrier  with height $|\langle V_{\rm AdS} \rangle| \sim m_{T_b}^2$ appears.
In order not to cause the decompactification of the modulus,
$H_{\rm inf} \sim \kappa \hat{M}^2 \lesssim m_{T_b}$
should be satisfied~\cite{Kallosh:2004yh, Conlon:2008cj},
where $H_{\rm inf}$ is the Hubble parameter during inflation. Unless the tadpole is extremely suppressed,
the inflation scale is bounded below as $H_{\rm inf} \gtrsim \oten{7}$\,GeV.\footnote{This is the case for $\zeta m_{3/2}/{\cal V} \gtrsim 1$\,GeV~\cite{Nakayama:2010xf}.}
Then ${\cal V}$ must be smaller than $\sim 10^7$, and the coefficient of the tadpole is  given by
\beq
 \frac{\zeta }{{\cal V}} m_{3/2} \;\gtrsim\; \zeta \times 10{\rm \,TeV}.
\eeq
In the case that there is an uplifting sector sequestered from the inflaton/visible sector as in the KKLT case,
one will find $\zeta \sim 1/\ln{\cal V} \gtrsim {\cal O}(10^{-2})$ without
a modification of Eq.(\ref{LVStadpole}).
Then only a slight tuning at ten percent level is necessary.
On the other hand, for $\zeta \sim 1$ in non-sequestered models, the tadpole is still
so large that the inflation scale is bounded below as $H_{\rm inf} \gtrsim \GEV{10}$ .
Therefore $\zeta$ must be suppressed by several orders of magnitude
with respect to the naive estimation.\footnote{This conclusion may be changed if the inflaton dynamics is significantly modified
by considering higher order terms in the K\"ahler potential. We leave this issue for future work.}

In the discussion above we have assumed $\omega_0 \sim {\cal O}(1)$.
Note that considering $\omega_0 \ll 1$ does not improve the tadpole problem,
in spite of the fact that this apparently suppresses the tadpole. This can be
understood as follows. The suppression of the tadpole in (20) is a remnant of
the no-scale structure, which is recovered in the large volume limit, ${\cal V} \rightarrow \infty$.
On the other hand, smaller $\omega_0$ reduces the mass of the overall volume modulus,
and therefore the ${\cal V}$ must be smaller in order not to destabilize the modulus for
a fixed inflation scale. Thus, the suppression of the tadpole becomes weaker: 
$\frac{\zeta m_{3/2}}{{\cal V}}/m_{T_b} \sim \zeta/{\cal V}^{1/2}$.

The overall volume modulus decays before big bang nucleosynthesis
if ${\cal V} \lesssim {\cal O}(10^{8})$, and the produced LSPs can account for the
dark matter if  ${\cal V} \sim 10^7$ and the LSP is Wino-like neutralino~\cite{Higaki:2012ar}.
Note that the cosmological moduli problem can be solved or relaxed by assuming
additional entropy production such as thermal inflation,\footnote{
One can easily implement the thermal inflation~\cite{Hindmarsh:2012wh} or second inflation~\cite{Nakayama:2012gh}
after the hybrid inflation in a unified manner.
} or the R-parity violation, and in this case, the upper bound on the volume will be relaxed.

So far we have focused on the tadpole of the inflaton. For successful inflation, not only the tadpole but also the mass
of the inflaton should be small enough. In general, there is so called $\eta$-problem in the $F$-term inflation in supergravity,
and some mild tuning of parameters is necessary to guarantee the light inflaton mass during inflation.
In the LVS model, there might be an effective quartic coupling of the inflaton $|\Phi|^4$ with a positive
coefficient in the K\"ahler potential,
which leads to a negative mass of order the Hubble parameter during inflation (see also Ref.~\cite{Buchmuller:2012ex}).
This can be cured by considering a coupling with the other moduli with an appropriate coefficient because these moduli have
an F-term whose size is larger than that of the inflaton.

%
%

%

\section{Discussion and Conclusions}
\label{sec:4}

So far we have assumed that the inflaton $\Phi$ is charged under the global U(1)$_\Phi$
symmetry, which is broken by the $M^2$-term. Here let us briefly discuss what if
there is no such global U(1)$_\Phi$ symmetry.
To be concrete, we consider the general case given in Sec.~2, and assume that $\Phi$ is
also singlet under U(1)$_R$ symmetry. Then, the inflation scale is naturally related to
the R-symmetry breaking. We expect naively $\kappa \sim m_{3/2}$ and $M \sim 1$,
but such large $M$ would result in too large tension of cosmic strings.
We may extend the gauge symmetry of the waterfall field to SU(2)$\times$U(1)
to avoid the cosmic string formation, which would allow larger values of $M$.
In any case,   the inflaton has a non-zero F-term of order $m_{3/2} M_P$,
 as the hybrid inflation and the Polonyi model are unified in some sense.
Interestingly, the inflation scale is tied to the SUSY breaking scale, $H_{\rm inf}
\sim m_{3/2}$, and the high-scale SUSY suggested by the SM-like
Higgs boson  may be simply due to the WMAP normalization of the
density perturbations. Although significant fine-tuning would be required to realize
the sufficiently small tadpole and the inflaton mass,
our claim that there must be a SUSY breaking singlet holds even in the absence of
the global U(1)$_\Phi$ symmetry.

The hybrid inflation is a simple and therefore attractive inflation model.
In the minimal supergravity, the  spectral index $n_s$ is predicted to be
$0.98 \hyp 1$, while it is possible to reduce $n_s$ slightly by adding a quartic
coupling of the inflaton in the K\"ahler potential~\cite{BasteroGil:2006cm}.
In fact, it is known that the spectral index close to unity is still allowed if there
is additional relativistic degrees of freedom coined ``dark radiation", the existence
of which is suggested by the current observations~\cite{Komatsu:2010fb,Dunkley:2010ge,Story:2012wx}.
Interestingly, the real component of the overall volume modulus generically decays into
its imaginary component, the axion~\cite{Higaki:2012ba}.
In the presence of the Giudice-Masiero term, the modulus decay can indeed produce
a right amount of dark radiation~\cite{Cicoli:2012aq,Higaki:2012ar}. Thus, in  light of the
tadpole problem and the tension of the predicted spectral index with observation, the LVS is an interesting
framework for hybrid inflation.

The inflation scale is constrained by the WMAP normalization of the density perturbation.
If the density perturbation is generated by some other mechanism such as the curvaton or
modulated reheating, the inflation scale can be lower, which relaxes the tension between the hybrid inflation and high-scale
SUSY.

Alternatively, it is also possible to build a low-scale inflation model
such as the two-field new inflation~\cite{Asaka:1999jb,Nakayama:2011ri} or alchemical inflation~\cite{Nakayama:2012gh}.
The inflation scale of the latter is necessarily lower than the gravitino mass, since
it utilizes the SUSY flat direction as the inflaton, which is lifted only by the soft SUSY breaking mass,
and therefore the moduli stabilization is not spoiled.

In this letter we have argued that the tadpole of the inflaton in the hybrid inflation, which causes a tension with high-scale SUSY,
can be cancelled if there is a SUSY breaking singlet with appropriate couplings with the inflaton. It is interesting that the presence
of such SUSY breaking singlet, which is not required  in high-scale SUSY from phenomenological point of view, is favored
by the inflation dynamics. We have also pointed out that the cancellation of the tadpole is automatic in a no-scale supergravity.
As an realistic set-up with no-scale structure, we have considered the LVS and shown that the tadpole is indeed suppressed
by the large volume, which enables us to implement the hybrid inflation in LVS
without severe fine-tuning.

\section*{Acknowledgment}
This work was supported by the Grant-in-Aid for Scientific Research on Innovative
Areas (No.24111702, No. 21111006, and No.23104008) [FT], Scientific Research (A)
(No. 22244030 and No.21244033 [FT]),
and JSPS Grant-in-Aid for Young Scientists (B) (No. 24740135) [FT].
This work was also supported by World Premier International Center Initiative (WPI
Program), MEXT, Japan, and by Grants-in-Aid for Scientific Research from the Ministry
of Education, Science, Sports, and Culture (MEXT), Japan (No. 23104008 and No.
23540283 [KSJ]).

\appendix

\section{Tadpole in the LVS}
\label{tadpolederi}

In this appendix, we will show the derivation of Eq.(\ref{tadpoler}).
For this purpose, we need to consider the moduli stabilization at first.
In the LVS model of Eq.(\ref{KWeff}), The scalar potential of moduli sector is given by
\bea
V_{\rm moduli} = \frac{|W|^2}{t^{3-n}f''}\bigg(
\bigg|
\frac{W_X}{W} - (n-1)\frac{f'}{t^n}
\bigg|^2 - n(n-1)\frac{f f''}{t^{2n}}
\bigg) \times \bigg(
1 + {\cal O}
\bigg(\frac{1}{t^n}\bigg)
\bigg).
\label{Vmoduli}
\eea
Here we defined $t \equiv T_b +T_b^{\dag}$, $x \equiv T_s +T_s^{\dag}$,
$W_X \equiv \partial_{T_s} W$ and $f' = \partial_x f$ and so on.
So long as $\kappa M^2 \Phi \ll 1$, one can neglect the presence of the inflaton for moduli stabilization.
The stationary condition is given by
\bea
\frac{W_X}{W} &=& \frac{1}{t^n} \bigg[
(n-1)f' - n \frac{xf''}{ax} + {\cal O}\left( \frac{1}{(ax)^2} \right)
\bigg],
\label{stationary1}
\\
f &=& \frac{2n}{3+n}\frac{xf'}{ax} + {\cal O}\left( \frac{1}{(ax)^2} \right) .
\label{stationary2}
\eea
Then, in the LVS model with $n=3/2$, one obtains $t^{n} \sim {\cal V} \sim e^{a \tau_s} \gg 1$ and $\tau_s^{3/2} \sim \xi$.
The vacuum energy density without any uplifting and the canonically normalized moduli masses are given by
\bea
\langle V_{\rm AdS} \rangle \sim  - \frac{1}{ax t^{3+n}} , \qquad
m_{T_b} \sim  \bigg(\frac{1}{ax t^{3+n}}\bigg)^{1/2}, \qquad m_{T_s} \sim  \frac{ax}{t^{3/2}} .
\eea
Next, the tadpole of $\Phi$ is estimated as
\bea
V_{\rm tadpole} &=&
\left[G_{\Phi} e^{G}(G_i G^i -3 ) + e^{G}( G^i \nabla_{\Phi} G_i + G_{\Phi}) \right]|_{\Phi =0} \Phi
\label{tadpoleVSUGRA_App}
\\
&=& r^* e^{K/2}m_{3/2}^* \kappa M^2 \Phi ,
\eea
where
\bea
m_{3/2} &=& e^{K/2} W \simeq \frac{\omega_0}{{\cal V}}, \\
\nonumber
r &=& -2 + \left(
K^{tt}K_t^2 +2 K^{tx}K_t K_x + K^{xx}K_x^2
\right) \\
\nonumber
&&
- K^{\bar{\Phi}\Phi}\left(
K^{tt}K_tK_{t\bar{\Phi}\Phi} + K^{tx}K_t K_{x \bar{\Phi} \Phi} +K^{tx}K_xK_{t \bar{\Phi} \Phi} +
K^{xx}K_xK_{x\bar{\Phi}\Phi}
\right) \\
&&+ \frac{W_X}{W}\left(
K^{tx}K_t + K^{xx}K_x - K^{\bar{\Phi}\Phi}(K^{tx}K_{t \bar{\Phi} \Phi} + K^{xx}K_{x\bar{\Phi}\Phi} )
\right).
\eea
After substituting the stationary conditions of Eq.(\ref{stationary1}) and (\ref{stationary2}) into the above expressions of $r$,
one can obtain Eq.(\ref{tadpoler}).
Alternatively, once one notes that the tadpole in the scalar potential is understood as
the relevant SUSY breaking term of $W= \kappa M^2 \Phi$,
more simple and general expression will be obtained as
\bea
r^* m_{3/2}^* = -2 F^{\varphi} - F^i \partial_i
\bigg[\ln \bigg(
\frac{\kappa M^2}{e^{-K/3}K_{\bar{\Phi} \Phi}}
\bigg)\bigg]
\eea
when the tadpole term of $\Phi$ is absent in the K\"ahler potential because of $U(1)_{\Phi}$.
Here $\varphi$ is the conformal compensator superfield and the SUSY breaking $F$-terms are given by
\bea
F^{\varphi} = m_{3/2}^* + \frac{1}{3}(\partial_i K) F^i,
\qquad
F^i = -e^{G/2}G^{i\bar{j}}G_{\bar{j}} .
\eea
Through this expression of $r$, one can expect
how the tadpole will be affected by the uplifting sector
and other moduli $T'$ generating the linear superpotential of $\Phi$ such that $\kappa M^2 = e^{-T'}$.


\begin{thebibliography}{99}

\bibitem{Komatsu:2010fb}
E.~Komatsu {\it et al.} [WMAP Collaboration],
Astrophys.\ J.\ Suppl.\ {\bf 192} (2011) 18
[arXiv:1001.4538 [astro-ph.CO]].

\bibitem{Guth:1980zm}
  A.~H.~Guth,
  Phys.\ Rev.\  {\bf D23}, 347-356 (1981);
A.~A.~Starobinsky,
Phys.\ Lett.\ B {\bf 91} (1980) 99;
  K.~Sato,
  Mon.\ Not.\ Roy.\ Astron.\ Soc.\  {\bf 195}, 467-479 (1981).


\bibitem{Copeland:1994vg}
  E.~J.~Copeland, A.~R.~Liddle, D.~H.~Lyth, E.~D.~Stewart and D.~Wands,
  Phys.\ Rev.\ D {\bf 49}, 6410 (1994)
  [astro-ph/9401011];
  G.~R.~Dvali, Q.~Shafi and R.~K.~Schaefer,
  Phys.\ Rev.\ Lett.\  {\bf 73}, 1886 (1994)
  [hep-ph/9406319];
  A.~D.~Linde and A.~Riotto,
  Phys.\ Rev.\ D {\bf 56}, 1841 (1997)
  [hep-ph/9703209].

\bibitem{Lazarides:1991wu}
  G.~Lazarides and Q.~Shafi,
  Phys.\ Lett.\ B {\bf 258}, 305 (1991).

\bibitem{Asaka:1999yd}
  T.~Asaka, K.~Hamaguchi, M.~Kawasaki, T.~Yanagida,
  Phys.\ Lett.\  {\bf B464}, 12-18 (1999)
  [hep-ph/9906366];
  Phys.\ Rev.\  {\bf D61}, 083512 (2000)
  [hep-ph/9907559].

\bibitem{Buchmuller:2000zm}
  W.~Buchmuller, L.~Covi and D.~Delepine,
  Phys.\ Lett.\ B {\bf 491}, 183 (2000)
  [hep-ph/0006168].

\bibitem{Nakayama:2010xf}
  K.~Nakayama, F.~Takahashi and T.~T.~Yanagida,
  JCAP {\bf 1012}, 010 (2010)
  [arXiv:1007.5152 [hep-ph]].

\bibitem{Buchmuller:2012ex}
  W.~Buchmuller, V.~Domcke and K.~Schmitz,
  arXiv:1210.4105 [hep-ph].


\bibitem{:2012gk}
  G.~Aad {\it et al.}  [ATLAS Collaboration],
  Phys.\ Lett.\ B {\bf 716}, 1 (2012)
  [arXiv:1207.7214 [hep-ex]].

\bibitem{:2012gu}
  S.~Chatrchyan {\it et al.}  [CMS Collaboration],
  Phys.\ Lett.\ B {\bf 716}, 30 (2012)
  [arXiv:1207.7235 [hep-ex]].



\bibitem{Okada:1990gg}
  Y.~Okada, M.~Yamaguchi and T.~Yanagida,
  Phys.\ Lett.\ B {\bf 262}, 54 (1991);\\
  see also
  Y.~Okada, M.~Yamaguchi and T.~Yanagida,
  Prog.\ Theor.\ Phys.\  {\bf 85}, 1 (1991);\\
  J.~R.~Ellis, G.~Ridolfi and F.~Zwirner,
  Phys.\ Lett.\ B {\bf 257}, 83 (1991);\\
  H.~E.~Haber and R.~Hempfling,
  Phys.\ Rev.\ Lett.\  {\bf 66}, 1815 (1991).

\bibitem{Giudice:2011cg}
  G.~F.~Giudice and A.~Strumia,
  Nucl.\ Phys.\ B {\bf 858}, 63 (2012)
  [arXiv:1108.6077 [hep-ph]];\\
  G.~Degrassi, S.~Di Vita, J.~Elias-Miro, J.~R.~Espinosa, G.~F.~Giudice, G.~Isidori and A.~Strumia,
  JHEP {\bf 1208}, 098 (2012)
  [arXiv:1205.6497 [hep-ph]];\\
see also
  F.~Bezrukov, M.~Y.~.Kalmykov, B.~A.~Kniehl and M.~Shaposhnikov,
  arXiv:1205.2893 [hep-ph].

\bibitem{Giudice:1998xp}
  G.~F.~Giudice, M.~A.~Luty, H.~Murayama and R.~Rattazzi,
  JHEP {\bf 9812}, 027 (1998)  [hep-ph/9810442].  


\bibitem{Randall:1998uk}
  L.~Randall and R.~Sundrum,
  Nucl.\ Phys.\ B {\bf 557}, 79 (1999)  [hep-th/9810155];  


\bibitem{Bagger:1999rd}
  J.~A.~Bagger, T.~Moroi and E.~Poppitz,
  JHEP {\bf 0004}, 009 (2000)  [hep-th/9911029].  

\bibitem{Kawasaki:2006gs}
  M.~Kawasaki, F.~Takahashi and T.~T.~Yanagida,
  Phys.\ Lett.\ B {\bf 638}, 8 (2006)
  [hep-ph/0603265];
  Phys.\ Rev.\ D {\bf 74}, 043519 (2006)
  [hep-ph/0605297].

\bibitem{Dine:2006ii}
  M.~Dine, R.~Kitano, A.~Morisse and Y.~Shirman,
  Phys.\ Rev.\ D {\bf 73}, 123518 (2006).






\bibitem{Endo:2006tf}
  M.~Endo, K.~Hamaguchi and F.~Takahashi,
  Phys.\ Rev.\ D {\bf 74}, 023531 (2006).




\bibitem{Asaka:2006bv}
  T.~Asaka, S.~Nakamura and M.~Yamaguchi,
  Phys.\ Rev.\ D {\bf 74}, 023520 (2006)
  [hep-ph/0604132].

\bibitem{Endo:2006qk}
  M.~Endo, M.~Kawasaki, F.~Takahashi and T.~T.~Yanagida,
  Phys.\ Lett.\ B {\bf 642}, 518 (2006)
  [hep-ph/0607170].


\bibitem{Endo:2007ih}
  M.~Endo, F.~Takahashi and T.~T.~Yanagida,
  Phys.\ Lett.\ B {\bf 658}, 236 (2008)
  [hep-ph/0701042].




\bibitem{Endo:2007sz}
  M.~Endo, F.~Takahashi and T.~T.~Yanagida,
  Phys.\ Rev.\ D {\bf 76}, 083509 (2007)
  [arXiv:0706.0986 [hep-ph]].



\bibitem{Coughlan:1983ci}
  G.~D.~Coughlan, W.~Fischler, E.~W.~Kolb, S.~Raby and G.~G.~Ross,
  Phys.\ Lett.\ B {\bf 131}, 59 (1983);
  A.~S.~Goncharov, A.~D.~Linde and M.~I.~Vysotsky,
  Phys.\ Lett.\ B {\bf 147}, 279 (1984);
  J.~R.~Ellis, D.~V.~Nanopoulos and M.~Quiros,
  Phys.\ Lett.\ B {\bf 174}, 176 (1986).


\bibitem{Jeong:2012en}
  K.~S.~Jeong and F.~Takahashi,
  arXiv:1210.4077 [hep-ph].

\bibitem{Cremmer:1983bf}
  E.~Cremmer, S.~Ferrara, C.~Kounnas and D.~V.~Nanopoulos,
  Phys.\ Lett.\ B {\bf 133}, 61 (1983).

\bibitem{Ellis:1983sf}
  J.~R.~Ellis, A.~B.~Lahanas, D.~V.~Nanopoulos and K.~Tamvakis,
  Phys.\ Lett.\ B {\bf 134}, 429 (1984);
  J.~R.~Ellis, C.~Kounnas and D.~V.~Nanopoulos,
  Nucl.\ Phys.\ B {\bf 247}, 373 (1984);
  J.~R.~Ellis, C.~Kounnas and D.~V.~Nanopoulos,
  Nucl.\ Phys.\ B {\bf 241}, 406 (1984).

\bibitem{Balasubramanian:2005zx}
  V.~Balasubramanian, P.~Berglund, J.~P.~Conlon and F.~Quevedo,
JHEP {\bf 0503}, 007 (2005)  [hep-th/0502058].

\bibitem{Higaki:2012ba}
  T.~Higaki, K.~Kamada and F.~Takahashi,
  JHEP {\bf 1209}, 043 (2012)
  [arXiv:1207.2771 [hep-ph]].


\bibitem{Cicoli:2012aq}
  M.~Cicoli, J.~P.~Conlon and F.~Quevedo,
  arXiv:1208.3562 [hep-ph].

\bibitem{Higaki:2012ar}
  T.~Higaki and F.~Takahashi,
  arXiv:1208.3563 [hep-ph].



\bibitem{Battye:2006pk}
  R.~A.~Battye, B.~Garbrecht and A.~Moss,
  JCAP {\bf 0609}, 007 (2006)
  [arXiv:astro-ph/0607339].
  R.~Battye, B.~Garbrecht and A.~Moss,
  Phys.\ Rev.\ D {\bf 81}, 123512 (2010)
  [arXiv:1001.0769 [astro-ph.CO]].

\bibitem{Dine:2010eb}
  M.~Dine, F.~Takahashi and T.~T.~Yanagida,
  JHEP {\bf 1007}, 003 (2010)
  [arXiv:1005.3613 [hep-th]].

\bibitem{Izawa:1996pk}
  K.~-I.~Izawa and T.~Yanagida,
  Prog.\ Theor.\ Phys.\  {\bf 95}, 829 (1996)
  [hep-th/9602180];
  K.~A.~Intriligator and S.~D.~Thomas,
  Nucl.\ Phys.\ B {\bf 473}, 121 (1996)
  [hep-th/9603158].


  \bibitem{Cline:1990bw}
  J.~M.~Cline and S.~Raby,
  Phys.\ Rev.\ D {\bf 43}, 1781 (1991);  
  R.~J.~Scherrer, J.~M.~Cline, S.~Raby and D.~Seckel,
  Phys.\ Rev.\ D {\bf 44}, 3760 (1991).  

\bibitem{Endo:2006xg}
  M.~Endo, K.~Kadota, K.~A.~Olive, F.~Takahashi and T.~T.~Yanagida,
  JCAP {\bf 0702}, 018 (2007)
  [hep-ph/0612263].



\bibitem{Ellis:1984bs}
  J.~R.~Ellis, C.~Kounnas and D.~V.~Nanopoulos,
  Phys.\ Lett.\ B {\bf 143}, 410 (1984).

\bibitem{Gelmini:1983ji}
  G.~Gelmini, C.~Kounnas and D.~V.~Nanopoulos,
  Nucl.\ Phys.\ B {\bf 250}, 177 (1985).

\bibitem{Ellis:2006ara}
  J.~R.~Ellis, Z.~Lalak, S.~Pokorski and K.~Turzynski,
  JCAP {\bf 0610}, 005 (2006)
  [hep-th/0606133].


\bibitem{Blumenhagen:2009gk}
  R.~Blumenhagen, J.~P.~Conlon, S.~Krippendorf, S.~Moster and F.~Quevedo,
JHEP {\bf 0909}, 007 (2009)
[arXiv:0906.3297 [hep-th]].

\bibitem{Becker:2002nn}
  K.~Becker, M.~Becker, M.~Haack and J.~Louis,
JHEP {\bf 0206}, 060 (2002)
[hep-th/0204254].

\bibitem{Shin:2011uk}
  C.~S.~Shin,
JHEP {\bf 1201}, 084 (2012)
[arXiv:1108.5740 [hep-ph]].

\bibitem{Conlon:2010ji}
  J.~P.~Conlon and F.~G.~Pedro,
JHEP {\bf 1006}, 082 (2010)
[arXiv:1003.0388 [hep-th]].

\bibitem{Choi:2010gm}
  K.~Choi, H.~P.~Nilles, C.~S.~Shin and M.~Trapletti,
JHEP {\bf 1102}, 047 (2011)
[arXiv:1011.0999 [hep-th]].




\bibitem{Cicoli:2012fh}
  M.~Cicoli, A.~Maharana, F.~Quevedo and C.~P.~Burgess,
JHEP {\bf 1206}, 011 (2012)
[arXiv:1203.1750 [hep-th]].

\bibitem{Kachru:2003aw}
  S.~Kachru, R.~Kallosh, A.~D.~Linde and S.~P.~Trivedi,
Phys.\ Rev.\ D {\bf 68}, 046005 (2003)
[hep-th/0301240].


\bibitem{Saltman:2004sn}
  A.~Saltman and E.~Silverstein,
JHEP {\bf 0411}, 066 (2004)  [hep-th/0402135];
%
  O.~Lebedev, H.~P.~Nilles and M.~Ratz,
Phys.\ Lett.\ B {\bf 636}, 126 (2006)  [hep-th/0603047];
%
  E.~Dudas, C.~Papineau and S.~Pokorski,
JHEP {\bf 0702}, 028 (2007)  [hep-th/0610297];
%
  H.~Abe, T.~Higaki, T.~Kobayashi and Y.~Omura,
Phys.\ Rev.\ D {\bf 75}, 025019 (2007)  [hep-th/0611024];
%
  R.~Kallosh and A.~D.~Linde,
 JHEP {\bf 0702}, 002 (2007)  [hep-th/0611183];
%
  H.~Abe, T.~Higaki and T.~Kobayashi,
Phys.\ Rev.\ D {\bf 76}, 105003 (2007)  [arXiv:0707.2671 [hep-th]].

\bibitem{Burgess:2003ic}
  C.~P.~Burgess, R.~Kallosh and F.~Quevedo,
JHEP {\bf 0310}, 056 (2003)
[hep-th/0309187];
%
  A.~Achucarro, B.~de Carlos, J.~A.~Casas and L.~Doplicher,
JHEP {\bf 0606}, 014 (2006)
[hep-th/0601190].

\bibitem{Kallosh:2004yh}
  R.~Kallosh and A.~D.~Linde,
JHEP {\bf 0412}, 004 (2004)
[hep-th/0411011].

\bibitem{Conlon:2008cj}
  J.~P.~Conlon, R.~Kallosh, A.~D.~Linde and F.~Quevedo,
JCAP {\bf 0809}, 011 (2008)
[arXiv:0806.0809 [hep-th]].

\bibitem{Hindmarsh:2012wh}
  M.~Hindmarsh and D.~R.~T.~Jones,
  arXiv:1203.6838 [hep-ph].

\bibitem{Nakayama:2012gh}
  K.~Nakayama and F.~Takahashi,
  arXiv:1206.3191 [hep-ph].

\bibitem{BasteroGil:2006cm}
  M.~Bastero-Gil, S.~F.~King and Q.~Shafi,
  Phys.\ Lett.\ B {\bf 651}, 345 (2007)
  [hep-ph/0604198].


\bibitem{Dunkley:2010ge}
  J.~Dunkley, R.~Hlozek, J.~Sievers, V.~Acquaviva, P.~A.~R.~Ade, P.~Aguirre, M.~Amiri, J.~W.~Appel {\it et al.},
  Astrophys.\ J.\  {\bf 739}, 52 (2011).
  [arXiv:1009.0866 [astro-ph.CO]].

\bibitem{Story:2012wx}
  K.~T.~Story, C.~L.~Reichardt, Z.~Hou, R.~Keisler, K.~A.~Aird, B.~A.~Benson, L.~E.~Bleem and J.~E.~Carlstrom {\it et al.},
  arXiv:1210.7231 [astro-ph.CO].



%
%
%
%
%
%
%
%
%
%
%
%
%
%
%
%
%
%
%
%
%
%
%

%
%
%
%
%
%


%
%

%
%
%
%
%
%
%
%
%
%
%
%
%
%
%
%
%
%
%
%
%
%
%
%
%
%
%
%
%
%
%
%
%
%
%
%
%
%
%
%
%
%
%
%
%
%
%
%
%
%

\bibitem{Asaka:1999jb}
  T.~Asaka, K.~Hamaguchi, M.~Kawasaki and T.~Yanagida,
  Phys.\ Rev.\ D {\bf 61}, 083512 (2000)
  [hep-ph/9907559];
  V.~N.~Senoguz and Q.~Shafi,
  Phys.\ Lett.\ B {\bf 596}, 8 (2004)
  [hep-ph/0403294].

\bibitem{Nakayama:2011ri}
  K.~Nakayama and F.~Takahashi,
  JCAP {\bf 1110}, 033 (2011)
  [arXiv:1108.0070 [hep-ph]];
  JCAP {\bf 1205}, 035 (2012)
  [arXiv:1203.0323 [hep-ph]].



\end{thebibliography}
\end{document}